\documentclass[%
twocolumn,
superscriptaddress,
amsmath,amssymb,
aps,
pre,
]{revtex4-2}

\usepackage{dcolumn}
\usepackage{bm}
\usepackage[mathlines]{lineno}

\usepackage[aboveskip=-10pt]{subcaption}
\usepackage{graphicx}
\usepackage{xcolor}

\begin{document}
	
	\title{Critical scaling and soft granular rheology of frictionless non-Brownian suspensions across jamming}
	
	\author{Rohan Vernekar}
	\email{vernekar.rohan@outlook.com}
	\affiliation{%
		Univ. Grenoble Alpes, CNRS, Grenoble INP, LRP, 38000 Grenoble, France
	}%
	\author{Romain Mari}%
	\affiliation{
		Univ. Grenoble Alpes, CNRS, LIPhy, 38000 Grenoble, France
	}%
	
	
	\author{Bruno Chareyre}
	\affiliation{
		Univ. Grenoble Alpes, Grenoble INP, 3SR, 38000 Grenoble, France
	}%
	
	\author{Hugues Bodiguel}
	\email{hugues.bodiguel@univ-grenoble-alpes.fr}
	\affiliation{%
		Univ. Grenoble Alpes, CNRS, Grenoble INP, LRP, 38000 Grenoble, France
	}%

	\date{\today}
	
	\newcommand{\exponent}{\mathcal{R}}
	\newcommand{\strainRateND}[1][]{%
		\def\temparg{#1}%
		\ifx\temparg\empty
		\dot{\gamma}^{\star}%
		\else
		(\dot{\gamma}^{\star})^{#1}%
		\fi
	}
	\newcommand{\Dphi}{\Delta \phi}
	
	\begin{abstract}
		We carry out 3D computer simulations to study suspension rheology across the jamming transition for deformable non-Brownian frictionless particles.
		The simulations are carried out at constant shear rates and suspension densities $(\strainRateND, \phi)$, for polydisperse spherical particles.
		The suspension is probed at densities in the neighbourhood of the jamming point $\phi_c$, as well as across much wider density ranges.
		We carry out critical scaling analysis of our close-to-jamming data and confirm that shear-driven jamming is a second-order critical phenomenon requiring strong corrections on shear.
		We build maps of the local strain rate exponent to classify flow regimes and examine the effect of the corrections on the suspension rheology.
		Further, using the critical scaling jamming density, we carry out soft granular rheological scaling over the wider density range, demonstrating an excellent data collapse.
		The soft granular rheology relations are able to predict our data over almost the entire dynamical and density range.
	\end{abstract}
	
	\maketitle
	
	\section{Introduction}
	\label{sec:intro}
	Particle suspensions play an important role in many physical and industrial systems.
	Cement slurries, blood flow, intestinal digesta, avalanches, mudslides etc.\ are some examples of particle laden suspension flows across varying length scales.
	When the typical particle sizes are $> 1~\mu m$, Brownian motion at the particle scale is negligible and such suspensions can be considered as athermal \cite{ness_PhysicsDenseSuspensions_2022, guazzelli_RheologyDenseGranular_2018}.
	If the suspended particles are deformable, and confining pressures sufficiently high, the suspension density or the particle packing fraction ($\phi$) can increase beyond the theoretical value set by the random close packing for rigid particles \cite{pan_ReviewShearJamming_2023a}.
	Such suspensions show complex physics that covers Newtonian, shear thinning and elastic disordered solid behaviours \cite{guazzelli_RheologyDenseGranular_2024}.
	
	The emergence of the yield stress in a soft suspension is called the jamming transition.
	This transition is seen when the suspension density goes above a critical volume fraction $\phi > \phi_c$ and the soft suspension turns into a disordered yield stress material, e.g.\ liquid foams \cite{bonn_YieldStressMaterials_2017}.
	In the rigid limit i.e.\ $\dot{\gamma} \to 0$, and under quasi-static conditions, $\phi_c$ defines the jamming point for an athermal suspension \cite{liu_JammingNotJust_1998}.
	This yield stress has been found to scale monotonically with increasing distance to the jamming point ($\Dphi = \phi - \phi_c$) as a power law $\tau_y \propto \Dphi^{y}$ \cite{heussinger_JammingTransitionProbed_2009}.
	While, when $\phi < \phi_c$ the suspension viscosity is known to diverge algebraically as $\eta \propto |\Dphi|^{-\alpha}$ \cite{boyer_UnifyingSuspensionGranular_2011}.
	
	This jamming transition from fluid to solid-like behaviours in athermal systems has been identified as a second-order critical phase transition \cite{olsson_CriticalScalingShear_2007, peshkov_UniversalityStressanisotropicStressisotropic_2022}, universal to shear and compression driven systems.
	The critical exponents at the jamming point determine both stress-divergence exponent below jamming and the yield stress exponent above jamming, which connects the diverging suspension viscosity below jamming, with yield-stress rheology above jamming for a driven soft suspension \cite{olsson_HerschelBulkleyShearingRheology_2012}.
	However, strong corrections to the critical scaling emerge even at small distances to $\phi_c$ and at small but finite strain rates \cite{olsson_CriticalScalingShearing_2011}.
	These corrections impair simple rheological scaling for the shear and normal stresses even at moderate density/strain-rate differences with the jamming point \cite{kawasaki_DivergingViscositySoft_2015}.
	
	When analysing data outside the immediate vicinity of the jamming point, a regime that more commonly reflects realistic conditions, \citet{kawasaki_DivergingViscositySoft_2015} introduced ``soft granular rheology'' (SGR) as an alternative framework for scaling stresses across the jamming transition.
	This approach is based on the physical idea that soft particle suspensions behave similarly to their hard particle counterparts, but with a renormalized particle diameter.
	SGR can be seen as an extension of the $\mu(J)$ granular rheology for rigid particle suspensions \cite{boyer_UnifyingSuspensionGranular_2011} to soft deformable particles.
	This approach has now also been extended to sheared systems in the inertial flow regime \cite{tapia_RheologySuspensionsNonBrownian_2024}.
	
	In this paper, we carry out 3D simulations over an extensive range of density values, covering the entire dense suspension regime, and over six decades of normalized strain rates, at constant $\phi$.
	We simulate polydisperse suspensions of frictionless particles using the discrete element method. Inter-particle interactions are determined by a lubrication model and harmonic repulsion \cite{chevremont_LubricatedContactModel_2020}.
	The details of the method are given in section~\ref{sec:method}, and the simulation parameters and shearing protocol in section~\ref{sec:simParams}.
	In the results section~\ref{sec:viscAcrossJamm}, we report the macroscopic measured quantities.
	We analyse our data using the critical scaling framework in section~\ref{sec:criticalScaling}, and SGR double power-law in section~\ref{sec:softGranRheo}.
	We present a conclusion in section~\ref{sec:conclusion}.
	
	\section{Method}
	\label{sec:method}
	We use the discrete element method (DEM) to simulate the motion of non-Brownian particles in a viscous liquid.
	For this purpose, we use the open source YADE-DEM solver \cite{angelidakis_YADEExtensibleFramework_2024, vaclavsmilauer_YadeDocumentation_2021}.
	The particles are spherical, polydisperse and considered to be made of two different materials --- a soft deformable particle core and a thin, highly rigid ``roughness'' shell around this core.
	The ratio of the shell thickness to the radius of the core is kept small $\varepsilon \ll 1$, and the ratio of their Young's moduli is set to a large value $E_{\varepsilon} / E  \gg 1$, in order to make the outer roughness shell essentially rigid.
	This model mimics asperities in suspended particles that prevent divergence of fluid lubrication forces upon particle contact and overlap.
	
	All particles are confined within a fully periodic box of size $L$. 
	On this box a uniform strain rate tensor $\bm{S}$ is imposed.
	The particle positions are updated according to the imposed strain rate tensor, and then corrected based on the sum of inter-particle forces experienced in one time step \cite{angelidakis_YADEExtensibleFramework_2024}.  
	The interaction forces come from two physical sources, steric contact between particles computed when the particles overlap and lubrication forces due to the interstitial fluid.
	In the following, we briefly describe the inter-particle interaction model. 
	For a detailed exposition of the visco-elastic coupling and the rheological model used, refer to  \cite{chevremont_LubricatedContactModel_2020}.
	This YADE-DEM solver with lubrication and roughness layer has been validated for various dense suspension flows, both with and without frictional contact between particles \cite{chevremont_NormalViscosityViscous_2024, chevremont_QuantitativeStudyRheology_2019}.
	
	\subsection{Contact interactions}
	We consider that all particles in the simulation have a layer of roughness, whose thickness is set by $\varepsilon a$, which regularizes the inter-particle interactions.
	Here, $a=(r_1+r_2)/2$ is the mean particle radius for two interacting particles with radii $r_1$ and $r_2$, respectively. 
	The deformation response for both the particle core and the roughness shell is linear with interaction force, but with differing material parameters.
	The normal contact force generated due to the overlap of the roughness layers of a pair of particles is given by $\bm{F}^c_n = -k_{\varepsilon}(\varepsilon a-u)\bm{\hat{n}}$, where $\bm{\hat{n}}$ is the unit vector along the centre-centre distance between the pair-particles.
	Here, $u$ is the gap between deformed particle surfaces, but excluding roughness layer.
	In the absence of overlap, $\bm{F}^c_n = 0$.
	
	If the deflection of the particle surface due to core deformation is given by $u_b$, the force-free configuration will have a gap $u_n = u + u_b$ between the particle centres.
	The gap $u$ is always positive, however both $u_n$ and $u_b$ can be positive or negative.
	Note that the normal deformation $u_b$ of the core material occurs due to contact as well as the normal lubrication force, and the particle surfaces can deflect towards each other due to attractive lubrication, giving $u_b > 0$.
	The particle-core material is also modelled as having a linear response to deformation, with a spring constant $k_b$.
	Thus, if the normal lubrication force is $\bm{F}^l_n$, the total normal force is given as $\bm{F}_n = \bm{F}^c_n + \bm{F}^l_n$, with $\bm{F}_n = k_b u_b \bm{\hat{n}}$.
	
	The tangential contact force is modelled assuming the Coulomb frictional interaction between particles.
	Therefore, $F^c_t \le \mu_0 F^c_n$, where $F^c_t$ is the magnitude of the contact tangential force, and $\mu_0$ is the microscopic friction coefficient between particles.
	The Coulomb criterion gives three regimes, no contact, sticking contact and slipping contact.
	If $\bm{v}_b$ is the elastic (deforming) contact, and $\bm{v}_t$ is the total tangential displacement, then the plastic slip is given by $\bm{v} = \bm{v}_t - \bm{v}_b$.
	The tangential contact force is given as $\bm{F}^c_t = \min(F^c_t,\, \mu_0 F^c_n)\bm{F}^c_t/|\bm{F}^c_t|$, where $-\bm{F}^c_t + \bm{F}^l_t = k_t \bm{v}_b$ is the linear response due to tangential contact and lubrication forces, where $k_t$ is the tangential material stiffness.
	
	\begin{figure*}
		\centering
		\includegraphics[width=\textwidth]{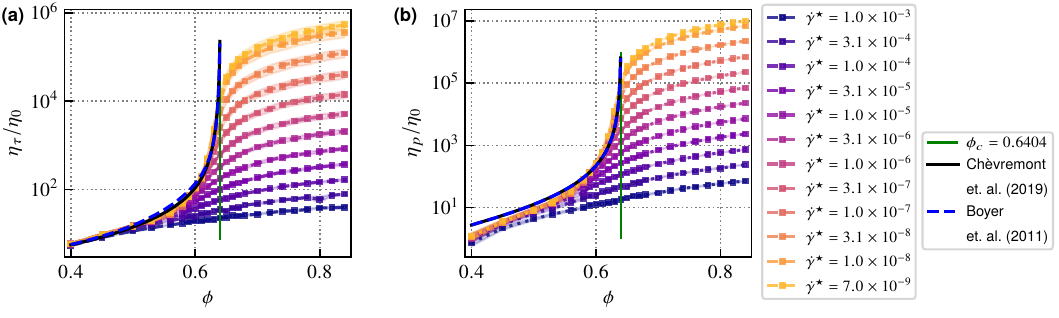}%
		\caption{Shear $\eta_{\tau}$ and normal viscosities across the jamming transition for various $\strainRateND$. The vertical (green) line shows the jamming density $\phi_c$. The curves below the jamming density show the divergence of viscosity for the rigid-particle rheology models from \citet{boyer_UnifyingSuspensionGranular_2011} (dashed, blue) and \citet{chevremont_QuantitativeStudyRheology_2019} (solid, black).
		}
		\label{fig:shearNormalVisc}
	\end{figure*}
	
	\begin{figure}
		\centering
		\includegraphics[width=0.499\textwidth]{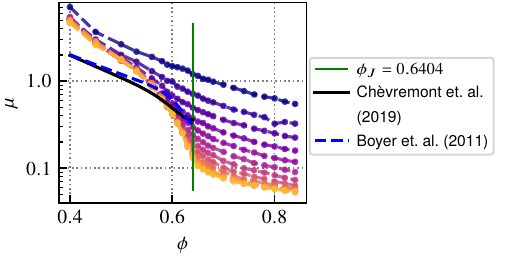}
		\caption{Friction coefficient plotted for various $\strainRateND$ values, crossing the jamming transition. The vertical line (green) shows the jamming density $\phi_c$. Curves below the jamming density are plotted for the macroscopic friction models of \citet{boyer_UnifyingSuspensionGranular_2011} (dashed, blue) and \citet{chevremont_QuantitativeStudyRheology_2019} (solid, black).
			The colour legend is the same as in figure~\ref{fig:shearNormalVisc}.
		}
		\label{fig:frictionCoeff_coarse}
	\end{figure}
	
	\subsection{Lubrication interactions}
	\label{subsec:methods:fluidModel}
	We use the normal and tangential lubrication force terms as in \citet{marzougui_MicroscopicOriginsShear_2015}, given as
	\begin{eqnarray}
		\label{eq:normalLubricationForce}
		\bm{F}^l_n &&= \frac{3}{2} \pi \eta_0 a^2 \frac{\dot{u}}{u} \hat{\bm{n}} \\
		\label{eq:tangentialLubricationForce}
		\bm{F}^l_t &&= \frac{\pi \eta_0}{2} \bigg( -2a + (2a+u) \ln\bigg(\frac{2a+u}{u}\bigg) \bigg) \dot{\bm{v}}
	\end{eqnarray}
	\noindent
	Here, $\eta_0$ is the suspension fluid viscosity, and the dotted quantities are the time rate of change of the gap.
	The linearity of the model allows for the analytical calculation of the discrete evolution of the normal gap using the particle positions at the previous time-step, when simple backward Euler discretisation of normal velocity, $\dot{u}$ is carried out.
	The tangential plastic slip $\bm{v}$ (or elastic displacement $\bm{v}_b$) can then be computed based on whether there is contact (none, stick, stick-slip) and the evaluation of the Coulomb criterion \cite{chevremont_LubricatedContactModel_2020}.
	Here again the tangential gap velocity evolution is discretised using the backward Euler scheme in time.
	
	\subsection{Motion integration}
	\label{subsec:methods:motionIntegration}
	The motion of individual particles is integrated in time using an explicit scheme \cite{vaclavsmilauer_YadeDocumentation_2021}.
	For a particle, the equations are written as
	\begin{equation}
		\label{eq:motionEquations}
		\frac{d}{dt} \begin{pmatrix}
			m \bm{\dot{r}} \\
			\bm{J} \bm{\Omega}
		\end{pmatrix} = \sum_i \begin{pmatrix}
			\bm{F} \\
			\bm{T}
		\end{pmatrix},
	\end{equation}
	\noindent
	where $m$ is the particle mass, $\bm{\dot{r}}$ the velocity, $J$ the moment of inertia tensor, $\bm{\Omega}$ the angular velocity, $\bm{F}$ and $\bm{T}$ are the forces and torques acting on the particle along the interaction links $i$. 
	This scheme gives a visco-elasto-plastic model for a soft suspension.
	
	\subsection{Stress calculation}
	\label{subsec:methods:stressCalc}
	We measure the instantaneous suspension stresses in the system as,
	\begin{eqnarray}
		\label{eq:stressCalc}
		\bm{\sigma} (t) = & V^{-1} \sum_{i>j} (\bm{r}^{(j)} - \bm{r}^{(i)}) \bm{F}^{l(ij)} \nonumber \\ &  + V^{-1} \sum_{i>j} (\bm{r}^{(j)} - \bm{r}^{(i)}) \bm{F}^{c(ij)}
	\end{eqnarray} 
	\noindent
	where the first term is the lubrication contribution to the total stress $\bm{\sigma}^l$ and the second term comes from the contact contribution $\bm{\sigma}^c$.
	The ${(ij)}^{\text{th}}$ component-wise suspension viscosity is given as $\eta_{ij} = \sigma_{ij} / \dot{\gamma}$.
	The contact and lubrication contributions to $\eta_{ij}$ can be computed from the respective stress components $\bm{\sigma}^c$ and $\bm{\sigma}^l$.
	
	\section{Simulation parameters}
	\label{sec:simParams}
	We carry out simulations at constant suspension densities $\phi$, and at constant simple shear strain rates $\strainRateND = \eta_0 S_{xy}/E$.
	In this study, we sweep non-dimensional shear rates over six decades $10^{-9} < \strainRateND < 10^{-3}$, with the suspension density in the dense regime.
	We carry out two sets of simulations, where the suspension density is varied over different ranges.
	A \textit{coarse} simulation set where $\phi$ is varied in the large range $0.4 < \phi < 0.84$, and a second \textit{fine} simulation set where the jamming transition is probed in the neighbourhood of the critical jamming density $\phi_c$, in the range $0.634 < \phi < 0.645$.
	In all simulations, the roughness layer is kept much smaller than the average particle size, and set as $\varepsilon = 10^{-3}$.
	Note that the density of the suspension is measured by including this layer in the diameter of the particle, $\phi = \frac{4\pi}{3L^3}\sum_i (a_i+\varepsilon a_i)^3$.
	
	We seek to understand and model the emergent suspension stress response under steady homogeneous conditions, for frictionless ($\mu_0=0$) soft particle suspensions.
	Since $k_{\varepsilon} \gg k_b$, the deformability of the soft particle core dominates the  dynamics, and we set $k_{\varepsilon} / k_b = 5\times 10^{4}$.
	The simulated particle sizes are polydisperse with a Gaussian polydispersity of 10\%.
	The suspension flow is studied in the Stokes flow regime without any inertial effects.
	We ensure that the simulations are in the over-damped limit by setting the particle Stokes number $St = \rho_0 a^2 S_{xy} / \eta_0 = 7\times 10^{-4}$, in all simulations by varying $\eta_0$ as the control parameter.
	
	\paragraph*{Shearing protocol:}
	We use 1000 particles for the \textit{coarse} simulation set, and 3000 particles for the \textit{fine} set \cite{kawasaki_DivergingViscositySoft_2015}.
	The particles are initialised at $\phi^{\mathrm{init}} = 0.1$ at pseudo-random positions. A uniform isotropic compressive strain given as $\dot{\gamma}_{xx,yy,zz} \propto |\phi - \phi^{\prime}|/\Delta t $ is then applied to the simulation box every iteration to reach the target density $\phi$, where $\phi^{\prime}$ is the density of the previous time step.
	Simultaneously, the suspension is also sheared with an increasing shearing strain rate given as  $\dot{\gamma}_{xy} = \dot{\gamma}_{xy} |\phi^{\prime} - \phi^{\mathrm{init}}| / (\phi - \phi^{\mathrm{init}})$.
	This is done in order to achieve better homogenisation, and avoid particle banding.
	Once the target suspension density is reached (with the criterion $|\phi - \phi^{\prime}| < 10^{-5}$) the compressive strain rate is switched off, and the suspension undergoes simple shearing strain rate of $\dot{\gamma}_{xy}$, until a total post-compression shear strain of $10$ is achieved.
	The steady state stresses and other macroscopic quantities for every parameter coordinate $(\strainRateND, \phi)$ are obtained by averaging over the last $7.5$ strains in each simulation  to avoid transient effects.
	
	\begin{figure*}
		\centering
		\includegraphics[width=\textwidth]{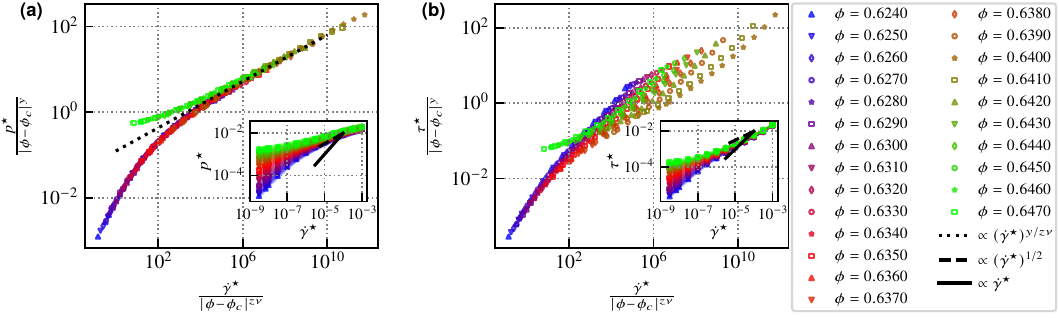}
		\caption{The critical scaling of the flow curves in the neighbourhood of the jamming point $\phi_c$ (\textit{fine} range), for the (a) pressure and (b) shear stress, without corrections to scaling.	
			The scaling is carried out with the critical exponents and the dotted line in (a) shows the shear-thinning strain rate exponent, at jamming $\phi = \phi_c$.
			The insets in both plots show the unscaled data.
		}
		\label{fig:criticalScalingRaw_fine}
	\end{figure*}
	
	\begin{figure*}
		\centering
		\includegraphics[width=\textwidth]{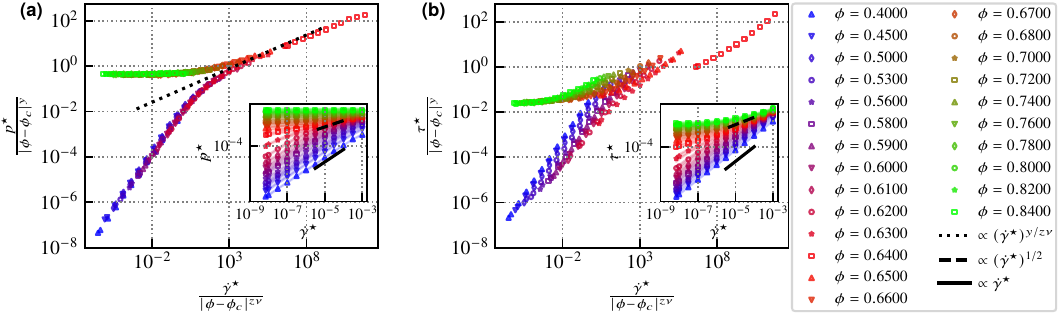}
		\caption{The critical scaling of the flow curves over the \textit{coarse} data range far away from jamming, for the (a) pressure and (b) shear stress, without corrections to scaling.
			The scaling is carried out with the critical exponents and the dotted line in (a) shows the shear-thinning strain rate exponent, at jamming $\phi = \phi_c$.
			The insets in both plots show the unscaled data.
		}
		\label{fig:criticalScalingRaw_coarse}
	\end{figure*}
	
	\section{Viscosities across jamming}
	\label{sec:viscAcrossJamm}
	The shear stress and pressure in the driven system are measured as, $\tau = (\sigma_{xy}+\sigma_{yx})/2$ and $p = -\mathrm{Tr}(\bm{\sigma})/3$, and normalized shear and normal suspension viscosities as, $\eta_{\tau} = \tau / (\eta_0\dot{\gamma})$ and $\eta_p = p / (\eta_0 \dot{\gamma})$.
	The data for shear and normal viscosities, for the \textit{coarse} simulation range are shown in figure~\ref{fig:shearNormalVisc} across the jamming transition, for various normalized strain rates.
	In the figure~\ref{fig:shearNormalVisc}, we also plot two rigid-suspension viscosity model curves, from \citet{boyer_UnifyingSuspensionGranular_2011} and \citet{chevremont_QuantitativeStudyRheology_2019}, for comparison with our data.
	In both models, the jamming density at which the viscosity diverges to infinity is taken as $\phi_c = 0.6404$ (shown by the vertical line), and no other parametric changes are made.
	
	The computation of the jamming $\phi_c$ in the rigid limit $\strainRateND \to 0$ is non-trivial and is described in detail in the next section~\ref{sec:criticalScaling}.
	The models appear to agree well with our data for both the shear and normal viscosities when the particle deformations are small, in the low strain rate regime $\strainRateND \lessapprox 3.1\times 10^{-8}$, below jamming $\phi_c$.
	The models slightly over-predict the normal viscosity at the lower end of the suspension density $\phi$ compared to our results.
	
	We also plot the macroscopic friction coefficient $\mu = \eta_{\tau} / \eta_p$ in figure~\ref{fig:frictionCoeff_coarse}, alongside the respective model predictions.
	Despite the seemingly good fit for the viscosity components, we see a marked mismatch between the model predictions and the friction data in the rigid limit $\strainRateND \to 0$, below jamming.
	This mismatch is largely due to the fact that the Boyer model finds the jamming friction coefficient, $\mu_c = \mu(\phi = \phi_c, \strainRateND \to 0) = 0.32$, while the Ch\`evremont model has taken this to be $0.36$, as a fitted parameter.
	In our case, we find that $\mu_c \approx 0.1$, in line with recent findings of \citet{kawasaki_DivergingViscositySoft_2015} and \citet{tapia_RheologySuspensionsNonBrownian_2024}.
	We think that this over-estimation of $\mu_c$ in these models likely comes because both these studies approach jamming only from the unjammed side, and never cross it.
	The other factor causing the mismatch is that our normal viscosity data is over-predicted by both the models in the lower density range.
	
	\section{Critical scaling analysis}
	\label{sec:criticalScaling}
	In order to accurately determine the jamming point $\phi_c$, we carry out critical scaling analysis using the \textit{fine} data-set ($0.634 < \phi < 0.645$).
	This analysis allows us to scale the quantities of interest (shear stress and pressure) using universal critical exponents at the jamming point.
	Using the scaling ansatz proposed by Olsson, Teitel and co-workers \cite{olsson_CriticalScalingShearing_2011, olsson_CriticalScalingShear_2007}, we can write any  quantity $\mathcal{P}$ as a function of the scaling variable $\Dphi / \strainRateND[1/z\nu]$ as,
	\begin{eqnarray}
		\label{eq:scalingAnstaz}
		\mathcal{P} (\Dphi, \strainRateND) &= \strainRateND[y_{\mathcal{P}}/z\nu] \left[ f_{\mathcal{P}} \left( \frac{\Dphi}{\strainRateND[1/z\nu]} \right) \right. \nonumber \\*
		& \left. + \strainRateND[\omega/z] g_{\mathcal{P}} \left( \frac{\Dphi}{\strainRateND[1/z\nu]} \right) \right],
	\end{eqnarray}
	\noindent
	where, $f_{\mathcal{P}}$ and $g_{\mathcal{P}}$ are scaling functions, $\nu$ is the correlation length exponent, $z$ is the dynamical critical exponent, $y_{\mathcal{P}}$ is the scaling dimension of quantity $\mathcal{P}$ (which in our case is the yield stress exponent for pressure or shear stress), and $\omega$ is the correction-to-scaling exponent.
	In the above scaling-ansatz, we have already incorporated the assumption that finite-size effects can be neglected for our simulation results.
	
	We approximate $f_{\mathcal{P}}$ and $g_{\mathcal{P}}$ to be exponential functions of fifth and third order polynomials in $x \equiv \Dphi / \strainRateND[1/z\nu]$, respectively.
	We simultaneously fit both shear stress and pressure data, using the least-squares Trust Region Reflective (TRF) algorithm.
	{For pressure, we restrict $g_{p}$ to a first-order polynomial exponent, as fits with higher-order polynomials exhibit severe parameter degeneracy.}
	The uncertainty weights of the data-points used for the minimized residuals are taken as the standard error of the mean (SEM).
	SEM is estimated by computing autocorrelation decay and the effective number of samples in the time-series data of the fitted quantities.
	
	Since the macroscopic friction coefficient remains finite and non-zero at $\phi_c$, we take $y=y_{\tau} = y_p$.
	We also make sure that our results are robust by fitting our data over progressively decreasing dynamic ranges of $\strainRateND < \strainRateND_{\max}$ \cite{olsson_CriticalScalingShearing_2011, olsson_DimensionalityViscosityExponent_2019}.
	The resulting $\chi^2/\text{DOF}$ plateaus at $\approx 1.7$, when $3.1 \times 10^{-6} \le \strainRateND_{\max} \le 3.1 \times 10^{-5}$.
	We take the fit values when $\strainRateND_{\max} = 3.1 \times 10^{-6}$ as the best-fit results.
	The errors on the fitted parameters are estimated using Jackknife resampling, and are reported as $\pm$ one standard deviation.
	
	From the fitting we obtain $\phi_c = 0.6404\pm 0.0005$, $y/z\nu = 0.270\pm 0.016$, $1/z\nu = 0.233\pm 0.012$, $\omega/z = 0.31\pm 0.018$, $y = 1.16\pm 0.02$ and $\alpha = 3.13\pm 0.23$.
	The $\phi_c$ value is in excellent agreement with the theoretical maximum random packing fraction for rigid spheres, and the critical exponents are also in excellent agreement with those reported in literature \cite{olsson_DimensionalityViscosityExponent_2019, degiuli_UnifiedTheoryInertial_2015, peshkov_UniversalityStressanisotropicStressisotropic_2022, tapia_RheologySuspensionsNonBrownian_2024}.
	
	We use the estimated exponents to scale the pressure and shear stresses measured from the \textit{fine} range simulations in figure~\ref{fig:criticalScalingRaw_fine}.
	Here we have ignored the corrections-to-scaling exponent $\omega$ when plotting for both pressure and shear stress for illustrative purposes.
	The scaling plots show an excellent double-branch data collapse for $p^{\star}$ but not for $\tau^{\star}$.
	This confirms that the corrections-to-scaling largely arise from the shear rheology, and have minimal effect on the pressure rheology \cite{olsson_CriticalScalingShearing_2011}.
	
	We further use the same critical exponents to rescale pressure and shear stresses measured from the \textit{coarse} data range in figure~\ref{fig:criticalScalingRaw_coarse}, again without the corrections-to-scaling.
	It is surprising to observe that the pressure data even at large distances from the jamming density collapses neatly onto a double branch.
	The shear data does not collapse, and shows systematic deviations from the double-branched collapse.
	
	\begin{figure*}
		\includegraphics[width=\textwidth]{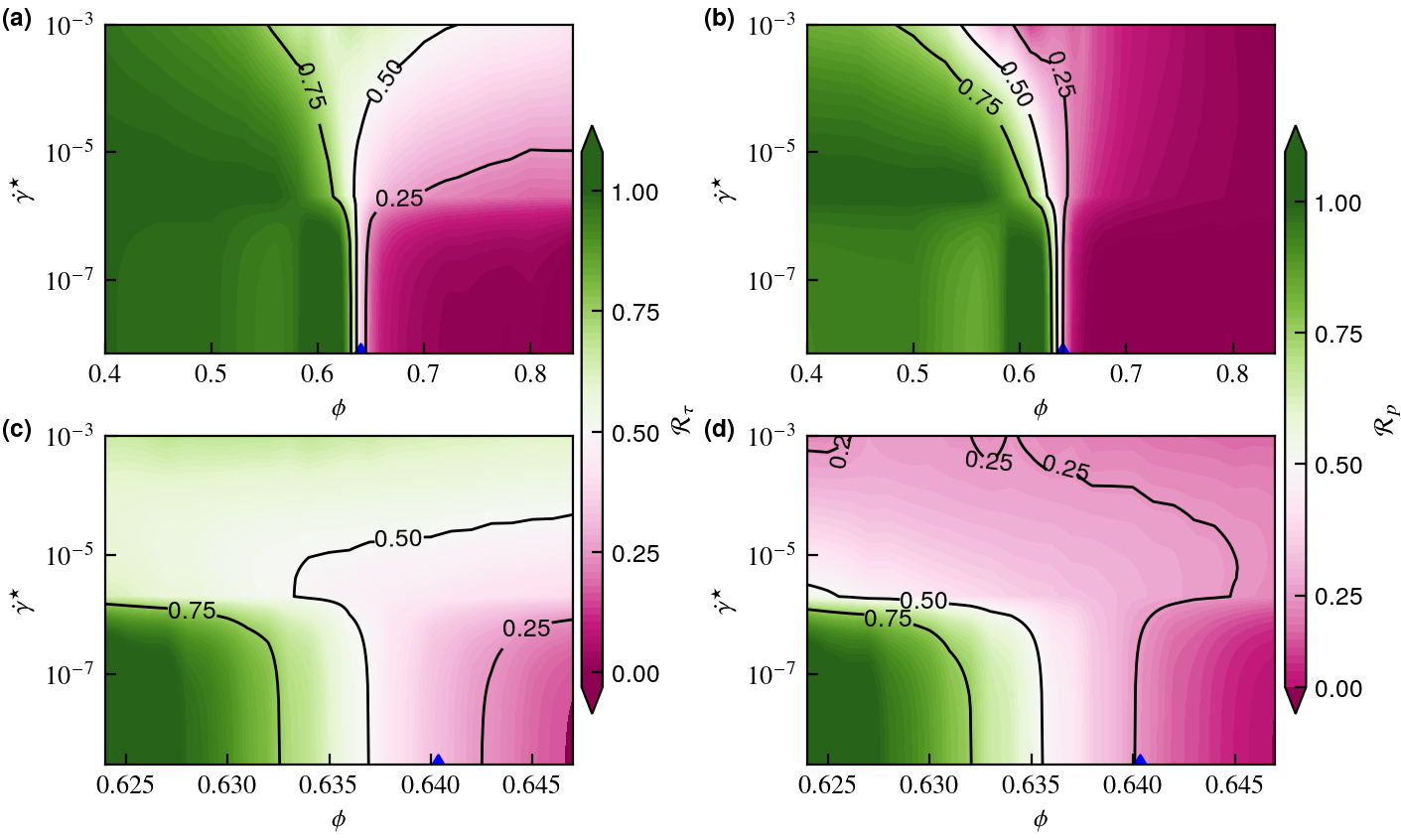}
		\caption{Flow regime maps that show the variation of the local exponent of the stress-strain rate measured from the flow curves in the log-log space. The panels (a) and (b) (top-row) show the variation for shear stress and pressure, respectively, over the \textit{coarse} $\phi$ range, while those in (c) and (d) show the corresponding values over the \textit{fine} $\phi$ range.  The plots show distinct transition zones from rate-independent viscosity regime $\mathcal{R} \to 1$, through the shear thinning regime $0.25 \alt \mathcal{R}  \alt 0.75$ to the yielding regime $\mathcal{R} \to 0$. The triangular marker (blue) indicates the jamming point $\phi_c$.}
		\label{fig:exponentMaps}
	\end{figure*}
	
	\section{Rate exponent maps}
	\label{sec:exponentMaps}
	
	Critical scaling analysis establishes that we require corrections to the critical scaling exponents as we move away from the jamming point $\phi_c$.
	In this section, we provide a rheological viewpoint on these corrections-to-scaling with increasing $\strainRateND$ and distance to jamming $|\Dphi|$.
	For this purpose we define local rate exponent for shear and pressure as $\mathcal{R}_{\tau}$ and $\mathcal{R}_{p}$, respectively.
	This implies, $\tau^{\star}(\phi, \strainRateND) \propto \strainRateND[\mathcal{R}_{\tau}]$ and $p^{\star}(\phi, \strainRateND) \propto \strainRateND[\mathcal{R}_{p}]$, locally, where the exponent $\mathcal{R}$ is also a function of $\Dphi$ and $\strainRateND$.
	
	{We compute this exponent $\mathcal{R}$ by fitting third order polynomials at constant $\phi$ to the shear and pressure flow curves in the log-log space, and computing the zeroth order term of its derivative.}
	These unscaled flow curves are shown in the insets of figures~\ref{fig:criticalScalingRaw_fine} and~\ref{fig:criticalScalingRaw_coarse}.
	We extract the exponent $\mathcal{R}$, by fitting for both the \textit{fine} and \textit{coarse} data-sets. 
	Note that an exponent of $\mathcal{R}=1$ means that the rheology is Newtonian, an exponent of $\mathcal{R}\approx0$ means that we are in the yield-stress dominant regime and an in-between value ($0.25 \alt \mathcal{R} \alt 0.75$) shows a transition crossing the shear-thinning regime.
	We plot exponent maps of $\mathcal{R}$ in figure~\ref{fig:exponentMaps}.
	The top row is for the \textit{coarse} data-set, for shear stress (left) and pressure (right).
	Similarly, the bottom row shows $\mathcal{R}$ variation for the \textit{fine} data-set, for shear stress (left) and pressure (right).
	
	In figure~\ref{fig:exponentMaps}(a) for the shear stress over the \textit{coarse} density range, we see three well-delineated regions, showing the Newtonian region (dark green) on the left, the yield-stress region (dark magenta) at the bottom-right and a shear-thinning region (magenta-white-green) at the top-right.
	Comparing with the $\mathcal{R}$ map for pressure in figure~\ref{fig:exponentMaps}(b), we see a marked difference in the aforementioned rheological regions.
	The shear-thinning region (approx. bounded by $0.25 \le \mathcal{R} \le 0.75$) becomes very narrow, and now presents a curvature opposite to that seen for the shear stress with increasing $\strainRateND$.
	The Newtonian region remains more or less similar, on the left of the map.
	While the yield-stress region on the right for pressure now extends to the highest $\strainRateND$ simulated.
	Near the jamming point $\phi_c, \strainRateND \to 0$ (shown with blue marker), the shear-thinning zone becomes very narrow for both $\tau^{\star}$ and $p^{\star}$.
	
	We plot similar exponent maps for the data in the neighbourhood of the jamming point (\textit{fine} range) in figures~\ref{fig:exponentMaps}(c) and (d) for shear stress and pressure, respectively.
	Below $\strainRateND \alt 10^{-6}$, the maps assume noticeable visual similarity.
	However, the location of the jamming point (triangular blue marker) shows that significant differences in rate exponent $\mathcal{R}$ persist in the neighbourhood of the jamming point.
	Importantly, comparisons also show that the deviation of the exponent, when moving away from $\phi_c$ along $|\Dphi|$ and $\strainRateND$, is markedly different for shear stress and pressure.
	This explains the difficulty encountered with estimating the jamming point and the rheological behaviour above and below jamming, when taking only a single stress value into account or approaching $\phi_c$ from a single direction.
	
	\begin{figure*}
		\centering
		\includegraphics[width=\textwidth]{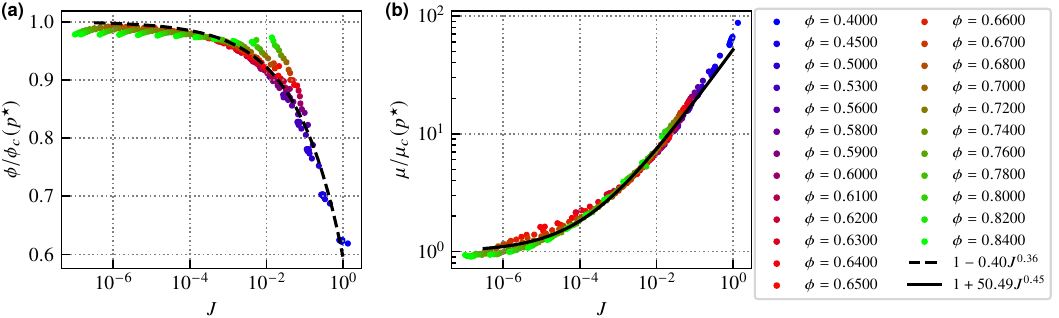}
		\caption{Plots show the rescaling of the rheological data with the soft granular rheological scaling. (a) shows the scaling of the suspension density, and (b) shows the scaling of the friction coefficient with the viscous number $J$. Overlaid curves show the fitted  $\phi(J,p^{\star})/\phi_c(p^{\star})$ law (dashed) and the $\mu(J,p^{\star})/\mu_c(p^{\star})$ law (solid), in (a) and (b) respectively.}
		\label{fig:softGranularRheo_coarse}
	\end{figure*}

	\begin{figure*}
		\centering
		\includegraphics[width=\textwidth]{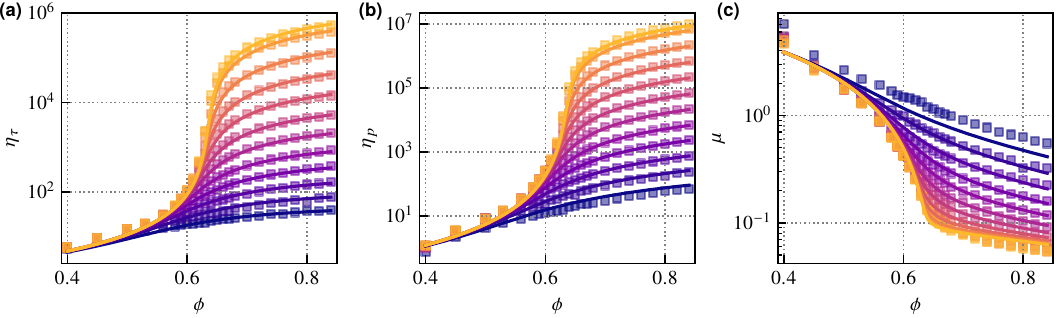}
		\caption{Plots show the predictions obtained from the `SGR' model (solid lines) overlaid on simulation data (square markers) for (a) normal viscosity, (b) shear viscosity and (c) macroscopic friction coefficient. The colour grading is the same as in figure~\ref{fig:shearNormalVisc}, indicating increasing non-dimensional strain rates ($\strainRateND$) when going from yellow to blue.}
		\label{fig:compareSGRpredictions}
	\end{figure*}
	\section{Soft granular rheology}
	\label{sec:softGranRheo}
	The ``soft granular rheology'' (SGR) approach was proposed as a means to scale the rheology of soft suspensions across the jamming transition \cite{kawasaki_DivergingViscositySoft_2015}.
	It is based on the simple physical idea that soft particle suspensions  behave similarly to their hard particle counterparts, but with a ``shifted'' volume fraction and friction coefficient.
	The idea can be considered as an extension of the hard particle suspension granular rheology (so-called $\mu(J)$ rheology) \cite{boyer_UnifyingSuspensionGranular_2011}, to soft particles.
	Soft particle compression, introduced by the confining pressure scale $p^{\star} = p/E$, leads to an effective change in the jamming points as a function of $p^{\star}$.
	Here, rheology is expressed through constitutive power laws of the viscous number $J=1/\eta_p = \eta_0 \dot{\gamma} / p$.
	In a similar vein, we term this as the $\mu(J,p^{\star})$ rheology, where the parameters $\phi$ and $\mu$ are given as,
	\begin{align}
		\label{eq:softRheo_mainPhi}
		\frac{\phi(J,p^{\star})}{\phi_c(p^{\star})} =   1 - a_{\phi} J^{b_{\phi}}, \\
		\label{eq:softRheo_mainMu}
		\frac{\mu(J,p^{\star})}{\mu_c(p^{\star})} = 1 + a_{\mu} J^{b_{\mu}}.
	\end{align}
	\noindent
	Here, $a_k$ and $b_k$ ($k\equiv \phi, \mu$) are the respective power-law parameters and $\phi_c(p^{\star})$ and $\mu_c(p^{\star})$ are the renormalization functions that ``shift'' the jamming volume fraction and friction coefficient, respectively. 
	
	The renormalization functions are also found as power-laws as,
	\begin{align}
		\label{eq:softRheo_shiftPhi}
		\frac{\phi_c(p^{\star})}{\phi_c(0)} = 1 + c_{\phi} (p^{\star})^{d_{\phi}} \\
		\label{eq:softRheo_shiftMu}
		\frac{\mu_c(p^{\star})}{\mu_c(0)} = 1 - c_{\mu} (p^{\star})^{d_{\mu}}.
	\end{align}
	\noindent
	where, $c_k$ and $d_k$ ($k\equiv \phi, \mu$) are the respective fitting parameters.
	
	We carry out $\mu(J,p^{\star})$ rheology over the \textit{coarse} range.
	We fix $\phi_c(0) = \phi_c = 0.6404$ as computed earlier from section~\ref{sec:criticalScaling}, and use least-squares (TRF) algorithm to obtain the power-law coefficients.
	For the $\phi(J, p^{\star})$ fit we get the parameters, $a_{\phi} = 0.404 \pm 0.008$, $b_{\phi} = 0.36\pm 0.01$, $c_{\phi} = 2.79\pm 0.18$ and $d_{\phi} = 0.79\pm 0.02$.
	For the $\mu(J, p^{\star})$ fit we obtain $\mu_c(0) = 0.083\pm 0.002$, $a_{\mu} = 50.5\pm 1.5$, $b_{\mu} = 0.447\pm 0.006$, $c_{\mu} = 1.662\pm 0.299$ and $d_{\mu} = 0.655\pm 0.068$.
	
	The collapsed suspension density and friction coefficient are shown in figure~\ref{fig:softGranularRheo_coarse}, overlaid with the modelled  equations~\ref{eq:softRheo_mainPhi} (dashed) and~\ref{eq:softRheo_mainMu} (solid) curves.
	We observe that systematic deviations are seen for the data-points that are farthest away from the jamming point along $|\Dphi|$.
	These deviations are much starker in the case of $\phi(J,p^{\star})$ than those for $\mu(J,p^{\star})$.
	Soft granular rheology has successfully been able to capture rheology across jamming in both simulations and experiments \cite{kawasaki_DivergingViscositySoft_2015, tapia_RheologySuspensionsNonBrownian_2024}.
	However, we note that in these studies the flow curves do not go deep into the yield-stress dominated flow regime, and remain in the shear-thinning region even above jamming.
	In our case, we cover a much wider suspension density range and a large dynamic range, yet the $\mu(J,p^{\star})$ scaling is able to capture the observed rheological behaviour reasonably well.
	
	Simple asymptotic analysis of the equations~\ref{eq:softRheo_mainPhi}--\ref{eq:softRheo_shiftMu} connects the power-law exponents with the critical jamming exponents estimated in section \ref{sec:criticalScaling} \cite{guazzelli_RheologyDenseGranular_2024}.
	We find that the yield stress emerges as a sum of two power-laws, $\sim|\Dphi|^{1/d_{\phi}}$ and $\sim|\Dphi|^{(1+d_\mu)/d_{\phi}}$.
	Hence, the yield stress critical exponent is given by the dominant asymptotic power-law as $y \approx 1/d_{\phi} = 1.266$.
	The crossover exponent is estimated as $z\nu = 1/d_\phi + 1/b_\phi \approx 4.04$, and the viscosity divergence exponent is estimated as $\alpha = 1/b_\phi \approx 2.78$.
	The shear-thinning exponent at the jamming point is obtained as $y/z\nu = b_{\phi} / (b_{\phi} + d_{\phi}) \approx 0.313$.
	These values obtained from SGR over the \textit{coarse} data-set are in reasonable agreement with the values obtained via rigorous critical scaling analysis in section~\ref{sec:criticalScaling}.
	
	We also carry out $\mu(J,p^{\star})$ rheological fitting of the data in the \textit{fine} range (plots not shown).
	From these closer-to-jamming fits, the yield stress critical exponent is $1/d_{\phi} \approx 1.105$, the crossover exponent is $1/d_{\phi} + 1/b_{\phi} \approx 4.60$, the viscosity divergence exponent $1/b_{\phi} \approx 3.496$ and the shear-thinning exponent at jamming point as $b_{\phi} / (b_{\phi} + d_{\phi}) \approx 0.24$.
	These values are in much closer agreement with the critical scaling exponents from section~\ref{sec:criticalScaling} than those estimated from the \textit{coarse} range.
	Note that here we obtain a much better SGR collapse for the \textit{fine} data than that over the \textit{coarse} range.
	
	We invert the SGR model to obtain a ``predictive'' relation for the normal viscosity as
	\begin{equation}
		\label{eq:normalViscSGR}
		\eta_p = \left[ \frac{a_\phi \left( 1 + c_\phi \eta_p^{d_\phi} \strainRateND[d_\phi] \right)}{1 + c_\phi \eta_p^{d_\phi} \strainRateND[d_{\phi}] - \phi / \phi_c(0)} \right]^{1/b_\phi}.
	\end{equation}
	\noindent
	From the known friction coefficient relation, we can then obtain the shear viscosity as $\eta_{\tau} = \mu(J,p^{\star}) \, \eta_p$ 
	Note that the relation obtained for $\eta_p$ is implicit and needs to be solved numerically.
	The SGR model predictions for shear and normal viscosities, and the friction coefficient are plotted in figure~\ref{fig:compareSGRpredictions}.
	We have good agreement for most of the curves for ($\eta_\tau, \eta_p, \mu$), except when the strain rate $\strainRateND \agt 3\times10^{-4}$, or when the density becomes very high $\phi \agt 0.8$. 
	Thus, the SGR model exhibits robust predictive capabilities for flows varying over large suspension densities and dynamic ranges.
	
	\section{Conclusion}
	\label{sec:conclusion}
	
	We have simulated the steady simple-shear rheology of frictionless non-Brownian suspensions of soft spherical particles, over the full dense regime and six decades of normalized strain rate $\strainRateND$.
	Close to jamming, a critical scaling analysis of the pressure and the shear stress locates the jamming transition at $\phi_c = 0.6404 \pm 0.0005$, in excellent agreement with random close packing of rigid spheres, and yields critical exponents $y/z\nu = 0.270\pm 0.016$, $1/z\nu = 0.233\pm 0.012$, $y = 1.16\pm 0.02$, $\alpha = 3.13\pm 0.23$ together with a sizeable correction-to-scaling exponent $\omega/z = 0.31\pm 0.018$.
	Remarkably, we find that the pressure adheres to the leading scaling form (without corrections), rescaling onto a double branch collapse, even at large distances from the jamming density, while the shear stress is strongly affected by the corrections, consistent with the anisotropic distortion of the contact network under shear. 
	Our maps of the local strain-rate exponent $\mathcal{R}$ make this asymmetry explicit in the
	$(\phi, \strainRateND)$ plane, delineating the Newtonian, shear-thinning and yield-stress-dominated regimes.
	The maps demonstrate that the various regime boundaries are traversed differently by the shear and normal stresses.
	
	Away from criticality, using the soft granular rheology $\mu(J, p^{\star})$ model, we obtain a reasonable collapse the rheology over the entire simulated range.
	In its asymptotic limits, the SGR model also recovers the critical exponents, most accurately when fitted to data close to $\phi_c$. 
	Physically, these results support a picture in which the contact-mediated elastic deformations endow the particles with a pressure-dependent effective size, so that a soft suspension behaves as a rigid-particle suspension whose jamming point drifts continuously with $p^{\star}$. 
	Using the SGR implicit relation in $\eta_p$ and $\eta_{\tau} = \mu(J, p^{\star}) \,\eta_p$, we can predict both viscosity components from $(\phi, \strainRateND)$ alone, going from the Newtonian regime deep into the yield-stress regime. 
	Extending this framework to frictional contacts is the natural next step.
	
	\bibliography{SoftHardSuspensionRheology_cited-only}

@article{ness_PhysicsDenseSuspensions_2022,
  title = {The {{Physics}} of {{Dense Suspensions}}},
  author = {Ness, Christopher and Seto, Ryohei and Mari, Romain},
  year = 2022,
  journal = {Annual Review of Condensed Matter Physics},
  volume = {13},
  number = {1},
  pages = {97--117},
  doi = {10.1146/annurev-conmatphys-031620-105938},
  urldate = {2022-12-08}
}

@article{guazzelli_RheologyDenseGranular_2018,
  title = {Rheology of Dense Granular Suspensions},
  author = {Guazzelli, {\'E}lisabeth and Pouliquen, Olivier},
  year = 2018,
  month = oct,
  journal = {Journal of Fluid Mechanics},
  volume = {852},
  publisher = {Cambridge University Press},
  issn = {0022-1120, 1469-7645},
  doi = {10.1017/jfm.2018.548},
  urldate = {2021-09-27},
  langid = {english}
}

@article{pan_ReviewShearJamming_2023a,
  title = {A Review on Shear Jamming},
  author = {Pan, Deng and Wang, Yinqiao and Yoshino, Hajime and Zhang, Jie and Jin, Yuliang},
  year = 2023,
  month = nov,
  journal = {Physics Reports},
  series = {A Review on Shear Jamming},
  volume = {1038},
  pages = {1--18},
  issn = {0370-1573},
  doi = {10.1016/j.physrep.2023.10.002},
  urldate = {2024-01-10}
}

@article{guazzelli_RheologyDenseGranular_2024,
  title = {Rheology of Dense Granular Suspensions across Flow Regimes},
  author = {Guazzelli, {\'E}lisabeth},
  year = 2024,
  month = sep,
  journal = {Physical Review Fluids},
  volume = {9},
  number = {9},
  pages = {090501},
  publisher = {American Physical Society},
  doi = {10.1103/PhysRevFluids.9.090501},
  urldate = {2024-12-20}
}

@article{bonn_YieldStressMaterials_2017,
  title = {Yield Stress Materials in Soft Condensed Matter},
  author = {Bonn, Daniel and Denn, Morton M. and Berthier, Ludovic and Divoux, Thibaut and Manneville, S{\'e}bastien},
  year = 2017,
  month = aug,
  journal = {Reviews of Modern Physics},
  volume = {89},
  number = {3},
  pages = {035005},
  publisher = {American Physical Society},
  doi = {10.1103/RevModPhys.89.035005},
  urldate = {2022-03-25}
}

@article{liu_JammingNotJust_1998,
  title = {Jamming Is Not Just Cool Any More},
  author = {Liu, Andrea J. and Nagel, Sidney R.},
  year = 1998,
  month = nov,
  journal = {Nature},
  volume = {396},
  number = {6706},
  pages = {21--22},
  publisher = {Nature Publishing Group},
  issn = {1476-4687},
  doi = {10.1038/23819},
  urldate = {2021-09-21},
  copyright = {1998 Macmillan Magazines Ltd.},
  langid = {english}
}

@article{heussinger_JammingTransitionProbed_2009,
  title = {Jamming {{Transition}} as {{Probed}} by {{Quasistatic Shear Flow}}},
  author = {Heussinger, Claus and Barrat, Jean-Louis},
  year = 2009,
  month = may,
  journal = {Physical Review Letters},
  volume = {102},
  number = {21},
  pages = {218303},
  publisher = {American Physical Society},
  doi = {10.1103/PhysRevLett.102.218303},
  urldate = {2022-03-26}
}

@article{boyer_UnifyingSuspensionGranular_2011,
  title = {Unifying {{Suspension}} and {{Granular Rheology}}},
  author = {Boyer, Fran{\c c}ois and Guazzelli, {\'E}lisabeth and Pouliquen, Olivier},
  year = 2011,
  month = oct,
  journal = {Physical Review Letters},
  volume = {107},
  number = {18},
  pages = {188301},
  publisher = {American Physical Society},
  doi = {10.1103/PhysRevLett.107.188301},
  urldate = {2021-09-24}
}

@article{olsson_CriticalScalingShear_2007,
  title = {Critical {{Scaling}} of {{Shear Viscosity}} at the {{Jamming Transition}}},
  author = {Olsson, Peter and Teitel, S.},
  year = 2007,
  month = oct,
  journal = {Physical Review Letters},
  volume = {99},
  number = {17},
  pages = {178001},
  publisher = {American Physical Society},
  doi = {10.1103/PhysRevLett.99.178001},
  urldate = {2022-01-03}
}

@article{peshkov_UniversalityStressanisotropicStressisotropic_2022,
  title = {Universality of Stress-Anisotropic and Stress-Isotropic Jamming of Frictionless Spheres in Three Dimensions: {{Uniaxial}} versus Isotropic Compression},
  shorttitle = {Universality of Stress-Anisotropic and Stress-Isotropic Jamming of Frictionless Spheres in Three Dimensions},
  author = {Peshkov, Anton and Teitel, S.},
  year = 2022,
  month = feb,
  journal = {Physical Review E},
  volume = {105},
  number = {2},
  pages = {024902},
  publisher = {American Physical Society},
  doi = {10.1103/PhysRevE.105.024902},
  urldate = {2023-07-20}
}

@article{olsson_HerschelBulkleyShearingRheology_2012,
  title = {Herschel-{{Bulkley Shearing Rheology Near}} the {{Athermal Jamming Transition}}},
  author = {Olsson, Peter and Teitel, S.},
  year = 2012,
  month = sep,
  journal = {Physical Review Letters},
  volume = {109},
  number = {10},
  pages = {108001},
  publisher = {American Physical Society},
  doi = {10.1103/PhysRevLett.109.108001},
  urldate = {2022-03-08}
}

@article{olsson_CriticalScalingShearing_2011,
  title = {Critical Scaling of Shearing Rheology at the Jamming Transition of Soft-Core Frictionless Disks},
  author = {Olsson, Peter and Teitel, S.},
  year = 2011,
  month = mar,
  journal = {Physical Review E},
  volume = {83},
  number = {3},
  pages = {030302},
  publisher = {American Physical Society},
  doi = {10.1103/PhysRevE.83.030302},
  urldate = {2022-03-08}
}

@article{kawasaki_DivergingViscositySoft_2015,
  title = {Diverging Viscosity and Soft Granular Rheology in Non-{{Brownian}} Suspensions},
  author = {Kawasaki, Takeshi and Coslovich, Daniele and Ikeda, Atsushi and Berthier, Ludovic},
  year = 2015,
  month = jan,
  journal = {Physical Review E},
  volume = {91},
  number = {1},
  pages = {012203},
  publisher = {American Physical Society},
  doi = {10.1103/PhysRevE.91.012203},
  urldate = {2021-09-19}
}

@article{tapia_RheologySuspensionsNonBrownian_2024,
  title = {Rheology of {{Suspensions}} of {{Non-Brownian Soft Spheres}} across the {{Jamming}} and {{Viscous-to-Inertial Transitions}}},
  author = {Tapia, Franco and Hong, Chong-Wei and Aussillous, Pascale and Guazzelli, {\'E}lisabeth},
  year = 2024,
  month = aug,
  journal = {Physical Review Letters},
  volume = {133},
  number = {8},
  pages = {088201},
  publisher = {American Physical Society},
  doi = {10.1103/PhysRevLett.133.088201},
  urldate = {2025-11-13}
}

@article{chevremont_LubricatedContactModel_2020,
  title = {Lubricated Contact Model for Numerical Simulations of Suspensions},
  author = {Ch{\`e}vremont, William and Bodiguel, Hugues and Chareyre, Bruno},
  year = 2020,
  month = jul,
  journal = {Powder Technology},
  volume = {372},
  pages = {600--610},
  issn = {0032-5910},
  doi = {10.1016/j.powtec.2020.06.001},
  urldate = {2021-06-18},
  langid = {english}
}

@article{angelidakis_YADEExtensibleFramework_2024,
  title = {{{YADE}} - {{An}} Extensible Framework for the Interactive Simulation of Multiscale, Multiphase, and Multiphysics Particulate Systems},
  author = {Angelidakis, Vasileios and Boschi, Katia and Brzezi{\'n}ski, Karol and Caulk, Robert A. and Chareyre, Bruno and Del Valle, Carlos Andr{\'e}s and Duriez, J{\'e}r{\^o}me and Gladky, Anton and Van Der Haven, Dingeman L.H. and Kozicki, Janek and Pekmezi, Gerald and Scholt{\`e}s, Luc and Thoeni, Klaus},
  year = 2024,
  month = nov,
  journal = {Computer Physics Communications},
  volume = {304},
  pages = {109293},
  issn = {00104655},
  doi = {10.1016/j.cpc.2024.109293},
  urldate = {2026-09-01},
  langid = {english}
}

@misc{vaclavsmilauer_YadeDocumentation_2021,
  title = {Yade Documentation},
  author = {{Vaclav Smilauer} and Angelidakis, Vasileios and Catalano, Emanuele and Caulk, Robert and Chareyre, Bruno and Ch{\`e}vremont, William and Dorofeenko, Sergei and Duriez, Jerome and Dyck, Nolan and Elias, Jan and Er, Burak and Eulitz, Alexander and Gladky, Anton and Guo, Ning and Jakob, Christian and {Francois Kneib} and Kozicki, Janek and Marzougui, Donia and Maurin, Raphael and Modenese, Chiara and Pekmezi, Gerald and Scholt{\`e}s, Luc and Sibille, Luc and Stransky, Jan and Sweijen, Thomas and Thoeni, Klaus and Yuan, Chao},
  year = 2021,
  month = nov,
  publisher = {Zenodo},
  doi = {10.5281/ZENODO.5705394},
  urldate = {2026-09-01},
  archiveprefix = {Zenodo},
  copyright = {Creative Commons Attribution 4.0 International, Open Access},
  langid = {english}
}

@misc{chevremont_NormalViscosityViscous_2024,
  title = {Normal Viscosity and {{Viscous}} Resuspension of Non-{{Brownian}} Suspensions},
  author = {Ch{\`e}vremont, William and Chareyre, Bruno and Bodiguel, Hugues},
  year = 2024,
  month = oct,
  number = {arXiv:2103.03718},
  eprint = {2103.03718},
  primaryclass = {cond-mat.soft},
  publisher = {arXiv},
  doi = {10.48550/arXiv.2103.03718},
  urldate = {2026-09-02},
  archiveprefix = {arXiv}
}

@article{chevremont_QuantitativeStudyRheology_2019,
  title = {Quantitative Study of the Rheology of Frictional Suspensions: {{Influence}} of Friction Coefficient in a Large Range of Viscous Numbers},
  shorttitle = {Quantitative Study of the Rheology of Frictional Suspensions},
  author = {Ch{\`e}vremont, William and Chareyre, Bruno and Bodiguel, Hugues},
  year = 2019,
  month = jun,
  journal = {Physical Review Fluids},
  volume = {4},
  number = {6},
  pages = {064302},
  publisher = {American Physical Society},
  doi = {10.1103/PhysRevFluids.4.064302},
  urldate = {2021-06-18}
}

@article{marzougui_MicroscopicOriginsShear_2015,
  title = {Microscopic Origins of Shear Stress in Dense Fluid--Grain Mixtures},
  author = {Marzougui, Donia and Chareyre, Bruno and Chauchat, Julien},
  year = 2015,
  month = jun,
  journal = {Granular Matter},
  volume = {17},
  number = {3},
  pages = {297--309},
  issn = {1434-7636},
  doi = {10.1007/s10035-015-0560-6},
  urldate = {2026-09-01},
  langid = {english}
}

@article{olsson_DimensionalityViscosityExponent_2019,
  title = {Dimensionality and {{Viscosity Exponent}} in {{Shear-driven Jamming}}},
  author = {Olsson, Peter},
  year = 2019,
  month = mar,
  journal = {Physical Review Letters},
  volume = {122},
  number = {10},
  pages = {108003},
  publisher = {American Physical Society},
  doi = {10.1103/PhysRevLett.122.108003},
  urldate = {2022-06-24}
}

@article{degiuli_UnifiedTheoryInertial_2015,
  title = {Unified Theory of Inertial Granular Flows and Non-{{Brownian}} Suspensions},
  author = {DeGiuli, E. and D{\"u}ring, G. and Lerner, E. and Wyart, M.},
  year = 2015,
  month = jun,
  journal = {Physical Review E},
  volume = {91},
  number = {6},
  pages = {062206},
  publisher = {American Physical Society},
  doi = {10.1103/PhysRevE.91.062206},
  urldate = {2022-03-09}
}
	
\end{document}